\documentclass[letterpaper]{article} 

\usepackage{aaai2026}

\usepackage{times}  
\usepackage{helvet}  
\usepackage{amsfonts}
\usepackage{courier}  
\usepackage[hyphens]{url}  
\usepackage{graphicx} 
\usepackage{amsmath}
\usepackage{amsthm}
\usepackage[justification=centering]{caption}
\usepackage{natbib}  
\usepackage{caption} 
\usepackage{algorithm}
\usepackage{algorithmic}

\usepackage{newfloat}
\usepackage{listings}
\DeclareCaptionStyle{ruled}{labelfont=normalfont,labelsep=colon,strut=off} 
\floatstyle{ruled}
\newfloat{listing}{tb}{lst}{}
\floatname{listing}{Listing}
\title{Expected Moral Shortfall for Ethical Competence\\in Decision-making Models}

\author{
    Aisha Aijaz\textsuperscript{\rm 1},
    Raghava Mutharaju\textsuperscript{\rm 2},
    Manohar Kumar\textsuperscript{\rm 3}
}
\affiliations{
    \textsuperscript{\rm 1}Dept. of Computer Science Engineering, IIIT-Delhi\\
    \textsuperscript{\rm 2}Mehta Family School for Data Science and Artificial Intelligence, IIT-Palakkad\\
    \textsuperscript{\rm 3}Dept. of Social Sciences and Humanities, IIIT-Delhi\\
    
}

\usepackage{bibentry}

\begin{document}

\maketitle

\begin{abstract}
Moral cognition is a crucial yet underexplored aspect of decision-making in AI models. Regardless of the application domain, it should be a consideration that allows for ethically aligned decision-making. This paper presents a multifaceted contribution to this research space. Firstly, a comparative analysis of techniques to instill ethical competence into AI models has been presented to gauge them on multiple performance metrics. Second, a novel mathematical discretization of morality and a demonstration of its real-life application have been conveyed and tested against other techniques on two datasets. This value is modeled as the risk of loss incurred by the least moral cases, or an Expected Moral Shortfall (EMS), which we direct the AI model to minimize in order to maximize its performance while retaining ethical competence. Lastly, the paper discusses the tradeoff between preliminary AI decision-making metrics such as model performance, complexity, and scale of ethical competence to recognize the true extent of practical social impact.
\end{abstract}


\section{Introduction}
AI systems have become increasingly ubiquitous in society, making crucial decisions that have implications at various degrees. Many efforts have been put into developing artificial cognition that may even surpass human intelligence in some cases, such as intellectual, emotional, physical, and mathematical \cite{cominelli2018seai,abdullah2022chatgpt,sheetz2020trends}. However, ethical competence, an understanding of which acts as one of the defining pillars of humanity, is far behind in terms of development and integration into AI systems \cite{brozek2019can}. 

Ethics is the study of right and wrong actions, and is defined normatively as a moral code that is acceptable to all rational people in society, with little regard to individual opinions \cite{sep-morality-definition}. This work aims to delve into the integration of ethical competence into decision-making AI models, based on its normative definition \cite{kagan2018normative}, thus involving a generalized and domain-applicable approach to artificial morality. In the course of our research, we did not come across an approach to making ethical competence in AI systems a more general cognitive ability.

Although there have been numerous contributions in the field of moral AI systems over the years, such as GenEth \cite{anderson2018geneth}, the Moral Machine Project \cite{awad2018moral}, the Consequence Engine \cite{vanderelst2018architecture}, and MoralDM \cite{dehghani2008integrated}, these are narrow in either their domain application, ethical theories used, or both. A reason why ethical competence is underexplored is because of the challenge it poses \cite{gordon2020building,cervantes2020toward}. Artificial decision-making on its own is a challenging task to accomplish, and this becomes increasingly complex with the addition of ethical judgment due to contradicting theories and inferred subjectivity \cite{bergmann2014challenges}.

Despite these challenges, there is a general consensus on the use of domain-specific ethical theories in order to resolve ethical issues \cite{holmes2018introduction}. Demarco \cite{demarco1997coherence} states that utilizing a view which coherently considers ethical principles, contextual judgement, and schools of thought would be the primary way to apply ethics to a decision. Most applications of modeling ethics in computer systems take from normative theories: consequentialism, virtue ethics, and deontology \cite{van2002deontic,greene2016embedding}. These may be applied to specific use cases, such as autonomous cars \cite{awad2018moral}, racial discrimination \cite{briggs2020mitigating}, and resource allocation \cite{10.1093/jlb/lsac012}.

Towards the intuition to ask whether AI systems can even be considered moral agents at all \cite{brozek2019can, formosa2021making}, the authors of this paper align with Whitby \cite{whitby2003myth}, who describes the morality of machines to be considered separate from human morality, in order to ever advance in the field of inherently ethical AI systems. Therefore, if an AI system can make critical decisions with ethical considerations effectively, it might as well be considered a moral agent.

Embedding moral constraints in existing decision-making AI models is imperative so as to not leave them to their own devices to hallucinate and ``always have an answer". It should be able to make ethically informed decisions while having a moral framework at the backend which does not solely comprise of a large corpus of general data with billions of parameters. Asking an LLM to provide ethical advice would do just that, however, it may not retain consistency or in some cases, even theoretical compatibility.

Developing AI systems with proper ethical competence would have a great impact on socially relevant spaces where they already readily exist, such as the approval of bank loans, the admittance of students to universities, distributing life-saving resources, and more.

To this end, this paper aims to bridge some of the gaps in the space of AI morality through the following contributions:

\begin{enumerate}
    \item A comparative analysis of techniques used to instill moral competence in learning models on the basis of classification metrics, ethical considerations, and model complexity.
    \item Expected Moral Shortfall (EMS), a metric derived from Expected Shortfall (ES) \cite{tasche2002expected}, a financial concept that, in our setting, minimizes expected moral failure beyond a certain threshold.
    \item An exploratory analysis of tradeoffs between classification metrics, the degree of incorporation of morality in decision-making, and the complexity of the models to provide alternatives that may best suit their purposes.
\end{enumerate}

We present the interdisciplinary terminology and mathematical notations used throughout the paper in the Preliminaries section. This is followed by an in-depth explanation of the Expected Moral Shortfall (EMS), and the methodology used. The experimentation and results for this work are presented, which apply the discussed theoretical concepts to two real-world datasets and examine the tradeoffs between various objectives of different AI models. Lastly, a brief exploration of the related works and discussion on this line of research and its societal impact is provided.   

\section{Preliminaries}
\textit{Ethically competent AI systems} are not to be confused with ethical AI systems. The prior involves morality as a layer of cognition for its decision-making process, while the latter involves ethically regulated development and deployment \cite{spiekermann2023value}. Ethically competent AI systems are what Moor defines as \textit{explicit ethical agents} which utilize one or more ethical theories to make decisions \cite{moor2006nature}.

Applying ethical competence to AI systems has its own challenges, however, the branch of applied ethics makes it simpler to access specific applicable theories in order to resolve dilemmas in a way that aligns with moral reasoning. Again, which theory might apply to which context may be subjective, however, there is consensus on which theories might give the best outcomes. \textit{Applied ethics} is a sub branch of ethics that thus deals with how ethical theories may be applied to real-world problems \cite{singer1986applied}. \textit{Normative ethics} is a parallel branch which recognizes the three broad theories under which all other ethical philosophies lie under some capacity \cite{kagan2018normative}. These involve consequentialism (ethics based on consequences), deontology (based on rights and duties), and virtue ethics (based on emulating virtuous behavior).  

Apart from theoretical aspects, there are also some pertinent contextual parameters that contribute significantly to a moral decision. One may argue that many seen and unseen parameters may play a role and sway decisions from one end of the moral spectrum to the other, however, in most cases, and for the sake of mathematical sobriety, we look at the few which count the most \cite{aijaz2025moralcompass}. With relation to consequentialism, we consider the quantification of \textit{severity}, \textit{utility}, and \textit{duration} of the consequence of an \textit{action}. For deontology we consider the \textit{moral intention} of the \textit{moral agent} doing the action. And for virtue ethics we consider the \textit{ethical principles} upheld and violated. Each of these are quantifiable based on the context and may be fine-tuned with weights to reflect certain ethical philosophies \cite{aijaz2025appleappliedethicsontology}. For example, a consequentialist approach would favor the characteristics of consequences more than the principles upheld or violated.

These values may be considered to determine an ethical judgement: \textit{morally right}, \textit{morally wrong}, and \textit{morally grey}. In order to discretize the range for morally grey actions, we may consider a context-sensistive threshold, as proposed by \cite{aijaz2025moralcompass}. This threshold would become narrower (harder constraints) when the domain of application is more established in terms of ethical guidelines, such as medicine, and would become wider (soft constraints) when the domain has fewer agreed upon guidelines such as the ethics of deploying AI systems in society.

An event $e$, that occurs in some universal set of events $E$ contains $n$ actions; $a$ belonging to set $A$. 
\begin{gather}
A(e) = \{{a_1, a_2, ..., a_n}\}, e \in E
\end{gather}

Next, we consider the weights $W$ associated with each normative theory as a tuple, where $\alpha$ is associated with consequentialism, $\beta$ with deontology, and $\gamma$ with virtue ethics. The flexibility provided by using such a tuple is great, as variations of these three broad theories may provide implications for a domain-specific ethical philosophy, $Ph$. 

\begin{equation}
W(Ph) = \{\alpha, \beta, \gamma\}\\   
\label{w}
\end{equation}

The event context $EC$ for an action $a$ may be defined as another tuple which, as discussed, are also associated with the normative schools of ethical thought. $C$ is the set of all consequences $c_i$ with their respective characteristics, $I$ represents deontology as moral intention which indicates the adherence to one's own duty and upholding the rights of others, and $P$ refers to the principles $pr$ which are upheld or violated. 
\begin{equation}
\begin{split}
    &EC = <C, I, Pr>\\
    &C(a) = {\{c_1, c_2, ...}\}\\
    &Pr(a) = {\{pr_1, pr_2, ...\}}\\
    &I(a) \in {\{i_1, i_2, ...\}}\\
\end{split}
\label{ec}
\end{equation}

The \textit{ethical judgement} $EJ$ for an action $a$ can be calculated using a weighted sum that takes a value from both (\ref{w}) and (\ref{ec}). Upon expansion we also see $g_i$, which would determine whether that particular term contributes positively or negatively to the judgement, based on the overall utility.  

\begin{equation}
\begin{split}
    &EJ(a) = W . EC\\
    &EJ(a) = \sum_{i=1}^{n} g_i . w_i . ec_i\\
\end{split}
\end{equation}

This value would determine a quantifiable \textit{ethics score} for action $a$, indicating a morally right action if it leans towards the positive end of the morality spectrum, and morally wrong if it leans towards the negative end. A value close to zero may be considered undetermined, or morally grey.

The ambiguity associated with a morally grey area may be discretized using a case-specific, context-sensitive thresholding mechanism \cite{aijaz2025moralcompass}. A \textit{default domain threshold} $\tau_d$ may be provided based on how strict one would want the constraints to be. This threshold may be adjusted via $\tau_{adjustment}$, however, sensitive to the contextual parameters considered in $EC$.

\begin{equation}
\label{thr}
    \tau \in \Bigl\{ \pm (\tau_{d}+ \sqrt{\frac{\sum_{i=1}^{n}(w_i.ec_i-\mu_{w.ec})^2}{n-1}}) \Bigl\}
\end{equation}

\begin{equation}
    \tau \in \Bigl\{ \pm (\tau_{d}+\tau_{adjustment} \Bigl\}
\end{equation}

Another consideration when making moral decisions is the prioritization of overall utility. This may be done either by striving towards higher values of $EJ$ or by minimizing the overall loss incurred by the morally worst-off cases. In literature, it is noted that akin to learning systems, ethicists too tend to prioritize avoiding catastrophic failures than encouraging supreme wins \cite{heath2014morality}. For this reason, we look at Expected Shortfall or $ES$, a risk measure that determines how bad a bad situation can actually get, overall. As a novel contribution, this paper applies a modified version of this core financial concept to the space of ethical competence in order to avoid morally high-risk situations using hard constraints. $ES$ is calculated as follows \cite{sowunmioptimization}, 

\begin{equation}
    ES_\theta(X) = -\frac{1}{\theta}(\mathbb{E}[X 1_{\{X\leq x_\theta\}}]+x_\theta(\theta - P[X\leq x_\theta]))
\end{equation}

where $x_\theta$ is the $\theta$-lower quantile of the bell-curve based on the chosen confidence interval, and $1_{\{X\leq x_\theta\}}$ is an indicator function that $\in{0,1}$ based on whether or not $x$ belongs to the subscript function. Thus, the $ES$ looks at the top $\theta \%$ of the worst-off cases (lowest $EJ(a_i)$) and targets those to avoid maximal moral risk. \cite{rockafellar2000optimization} propose a convex function to minimize $ES$ with some loss function $f(\omega,r)$ and a threshold $t$ as,

\begin{equation}
    F_\theta(\omega, t) = t + \frac{1}{1-\theta}\mathbb{E}[(f(\omega,r)-t)^+] 
\end{equation}

Since this function is convex and continuously differentiable, the risk or overall \textit{loss} associated with any variable $\omega$ may be determined by minimizing $F_\theta$. 

\section{Expected Moral Shortfall}
Combining the information presented above, we use the discretization of abstract ethical concepts such as consequences and intentions along with the presented context-sensitive threshold values post-adjustment, to propose an \textit{Expected Moral Shortfall} (EMS). This would serve as an added term to an existing loss function that may be embedded into the decision capacity of a learning model such as a deep neural network (DNN). Doing so would provide an apparent ethical competence to the AI model due to its moral consideration and selective hard moral constraints. 

Given the ethical judgement $EJ$ and a context-sensitive threshold $\tau$, the $EMS$ may be given as follows,

\begin{equation}
    EMS = \min_{\tau}[\tau + \frac{1}{1-\theta}.\frac{1}{n}\sum\max(\tau-EJ(a_i),0)]
\label{ems}
\end{equation}

We expect a decline in model performance with the inclusion of moral consideration. However, using EMS, we can flexibly decide how large $\theta$ can be, which would subsequently decide the percentage of the top morally-worst cases. A large $\theta$ would affect a larger number of cases on which the hard constraint of $\tau$ would be applied, whereas a smaller $\theta$ would allow some moral consideration to most high risk cases, further reducing its impact on model performance. During our dataset processing and model experimentation we realized that most cases are closer to the normal in terms of $EJ$, which is expected as recorded items in nature tend to follow the bell curve \cite{krithikadatta2014normal}. Thus, by making the shortfall cutoff flexible based on the criticality of the domain, one may adjust the level of ethical competence of the model as per their needs in order to retain model performance.

\section{Methodology}
To describe our method we have identified two real-world datasets which have been discussed in detail below. Additionally, we consider learning models of various type and complexity to demonstrate the applicability of ethical competence into such systems. Finally, we compare the chosen objectives for these models against the proposed EMS loss minimization task.

\subsection{Datasets}
In order to demonstrate the integration of ethical competence into a learning model, we have used two datasets from different domains: a graduate admissions dataset that includes various features about university applicants and their admission status, and a loan approval dataset that includes features of loan applicants and their application status.

Using the philosophical metrics described above, we can use our own discretion to recognize which features would contribute to a moral decision and how. These features are developed manually, thus incorporating an aspect of human-in-the-loop. Although we have LLMs and other AI models that may provide us with ethical feature mappings for these datasets, they may not be consistent or explicable. This is why it is imperative that these features be created using human oversight\footnote{The authors would also like to clarify that these feature mappings are not absolute for this domain. Perhaps another user may map these features differently, and that would reflect differently in the learning models' performance. This would not be \textit{wrong}, rather an \textit{alternative} way to align moral values to real-world, domain-specific concepts. Our approach aims to demonstrate flexibility to the user based on their own preferences, owing to subjectivity and differences in moral thought. This is why these features and the values for normative weights may be adjusted flexibly.}. The learning may then be done over the datasets using AI models.

\subsubsection{Dataset 1: Graduate Admissions Dataset}
The graduate admissions dataset \cite{gradadmissionsdataset} consists of standard metrics that are required of an applicant when applying for higher education such as their GRE score \cite{bleske2014trends}, English language (TOEFL) score \cite{wait2009relationship}, the rating of the university they are applying to, whether or not they have some research experience, and the strength of their Statement of Purpose (SOP), and Letters of Recommendation (LORs) \cite{zhang2025research}. The dataset itself is small-scale and was partially augmented to improve its target feature to be more balanced. It thus had the graduate application information of 770 applicants. These features were encoded and normalized to provide a clean processed dataset for feature mapping.

We had to identify which features would contribute to the normative ethics weights and in what capacity. This required the determination of nine new features, of which three corresponded to the normative ethics weights: $\alpha$, $\beta$, $\gamma$ (See equation (\ref{w})); and six corresponded to their contextual values: $C_{severity}$, $C_{utility}$, $C_{duration}$, $Pr_{upheld}$, $Pr_{violated}$, and $I$ (See equation (\ref{ec})). 

The values for $W$ can be chosen as per the user's flexible needs. If the user would like to guide the model to learn consequentialist outcomes, the $W$ tuple with values for $\alpha$, $\beta$, and $\gamma$ would look something like $<0.8, 0.1, 0.1>$. On the other hand, if the user chooses a philosophy that is more specific to the domain, such as healthcare, they may pick something from the bioethics space such as Principlism, for which the $W$ tuple could look like $<0.3, 0.6, 0.1>$. In this way, the broader theories can be adjusted to emulate established ethical theories.

For $EC$, the following features were mapped from the existing applicant details, $\forall{i \in N}$, where $M$ is the overall score of the student based on their GRE, CGPA, and TOEFL, and $N$ is the number of instances in the dataset,
\begin{equation}
\begin{split}
    &C_{severity} = \mu (gre_i,cgpa_i)\\
    &C_{utility} = \begin{cases}\frac{M_i}{penalty_{high}}~&{\text{ if }}~M_i \leq pass~,\\\frac{M_i}{penalty_{low}}~&{\text{ if }}~M_i > pass~.\end{cases}\\
    &C_{duration} =\begin{cases}1.\mu(uni\_rank_i, cgpa_i)~&{\text{ if }}~res_i = 1~,\\0.5.\mu(uni\_rank_i, cgpa_i)~&{\text{ if }}~res_i = 0~.\end{cases}\\
    &Pr_{upheld} = \mu (gre_i,cgpa_i, sop_i, lor_i)\\
    &Pr_{violated} = \begin{cases}
    M_i&{\text{if }}M_i > pass,~adm_i < 0.5,\\
    0&{\text{if}}~M_i \leq pass,~adm_i \geq 0.5~.\end{cases}\\
    &I = \mu(sop_i, research_i)\\
\end{split}
\end{equation}

The greater the scores of the student, the greater the severity of rejection. Similarly, if the student has high overall scores, rejection would cost a high penalty, thus lowering the utility of the action, and vice versa. If the student has previous research experience, the university can assume with higher probability that the duration of a rejected consequence would be longer as well. If the overall scores, SOP, and LOR ratings are high, acceptance would uphold principles of fairness and meritocracy. Whereas, if the student has passing scores, where $pass$ indicates the universities standard for admittance, but has been rejected admittance ($adm_i<0.5$), this would be violation of these principles. Finally, the intention of the potential student may be gauged by the quality of their SOP and LORs.

With the addition of these nine new \textit{moral} features, we were then able to calculate values for the ethical judgement $EJ$ and the context-sensitive thresholds $\tau$ for each instance in the dataset. Once these were calculated, finding the new status of graduate admissions was only a matter of threshold-based decisions, akin to ReLU, where if the value of $EJ$ was below the threshold $\tau$, the student's application was annotated as 0, and otherwise 1. For the purpose of model experimentation, two versions of the datasets were retained: the original with no moral inclination, and one with the morally inclined decisions as the target feature and all other intermediate columns ($EC$, $W$) dropped. 

\subsubsection{Dataset 2: Loan Approval Dataset}
Similar preprocessing techniques were applied to the loan approval dataset \cite{loanapprovaldataset}, as done for Dataset 1. The dataset is a large-scale one with 11 feature columns and a target column for the loan application status (accept, reject). It consisted of 32,582 fairly balanced data points, due to which no further augmentation or preprocessing was required. 

The features for this dataset included the applicant's age, income, home ownership status, employment length, loan intent, loan grade, loan amount, and interest rate. Additionally, columns are provided for the person's credit history and whether or not they have previously defaulted. Finally, the target column indicates whether or not their loan request is accepted or rejected. These existing features were mapped to the ethical concepts, $\forall{i \in N}$, using manual intervention, replicating the process done in Dataset 1, however, specific to this dataset. 
\begin{equation}
\begin{split}
    &C_{severity} = \frac{loan\_amt}{person\_income}\\
    &C_{utility} = \frac{loan\_intent_i.loan\_amt_i}{\max(LOAN\_AMT)}\\
    &C_{duration} = \min(1,(\frac{1}{1+emp\_length}+\frac{1}{1+cred\_hist})) \\
    &Pr_{upheld} = \begin{cases}1~&{\text{ if }}~(low/high)\_income,~!defaulter,\\0.5~&{\text{ if }}~high\_income,~defaulter,\\0~&{\text{ if }}~low\_income,~defaulter.\end{cases}\\
    &Pr_{violated} = \max(defaulter, loan\_grade)\\
    &I = loan\_int\\
\end{split}
\end{equation}

If the loan amount is high and the income of the person is low, this may indicate to the bank that the applicant is taking a risky loan, rejecting which may cause severe consequences for them. The utility factor also takes into account the intention of the loan, increasing utility if the loan amount is high and taken for the purpose of education or medical needs, and reduced if taken for personal ventures or home improvements. The duration of the consequence is inversely proportional to the length of employment of the applicant and their credit history, given the fact that the longer a person has been employed, the more option they may have for recovery if their loan request is rejected. This implies that newer employees may suffer longer if their request is rejected. The principles of fiduciary responsibility and fairness may be upheld or violated based on the decisions considering the applicant's income, defaulter status, and the grade of their loan. Finally, the moral intention of the applicant may be gauged by the intention for applying for a loan. These values which are a mix of categorical and continuous were all encoded, normalized, and preprocessed to provide the moral features, corresponding $EJ$ and $\tau$ columns, and finally the updated loan status target column with moral consideration. 

\subsection{Ethical Competence in Models of Varying Complexity and Design}
The next step was to determine methods of embedding morality in learning models. We identified three main ways:
\begin{enumerate}
    \item Introducing a moral penalty in the learning weights,
    \item Overriding decisions post-hoc, and
    \item Modifying the loss function of the learning model.
\end{enumerate}
These techniques, where feasible, could be used for gauging their applicability to models of different types and complexities. In order to test our EMS model, we wanted to do a robust analysis of tradeoffs in objectives using these three techniques and for different classification models:
\begin{enumerate}
    \item Logistic Regression (Linear function),
    \item Naive Bayes (Probabilistic function),
    \item Random Forest (Tree Ensemble),
    \item Support Vector Machine (Margin-based), and
    \item Deep Neural Network (Deep learning architecture).
\end{enumerate}
The third consideration we made when determining ethical competence in AI models was the optimization of various important objectives. There were:
\begin{enumerate}
    \item Classification Performance,
    \item Model Complexity, and
    \item Ethical Competence.
\end{enumerate}
We tested our proposed EMS model against these different baselines to provide an in-depth analysis on how ethical competence may be incorporated into an AI model based on the needs of the engineer or demands of the use case. These results are discussed in the following Experiments section.

\subsection{EMS Loss Minimization}
Using equation (\ref{ems}), we can determine the Expected Moral Shortfall or in terms of machine learning, the loss $\mathbb{L}_{EMS}$ that may be applied to an applicable deep learning model. Since we received the highest baseline results on the DNN model, we applied the weighted loss to the model's original loss function, which was binary cross-entropy \cite{ruby2020binary}.

\begin{equation*}
    \mathbb{L}_{EMS}(X) = \min_{\tau}[\tau + \frac{1}{1-\theta}.\frac{1}{n}\sum\max(\tau-EJ(a_i),0)]
\end{equation*}
\begin{equation}
    \mathbb{L}_{BCE}(X) = -\frac{1}{N}\sum_{i=1}^{N}[y_i\log\hat{y_i}+(1-y_i)\log(1-\hat{y_i})]
\end{equation}
\begin{equation}
    \mathcal{L} = \mathbb{L}_{BCE}(X) + \lambda.\mathbb{L}_{EMS}(X)
\end{equation}

The loss function of the EMS model $\mathcal{L}$ may be then adjusted using the $\lambda$ coefficient to manipulate how much the EMS would affect overall loss. 
\section{Experiments and Tradeoffs Analysis}

In order to gather comprehensive empirical results and to take the embedding of morality into learning systems one step further, we ran different models on a combination of the original and modified datasets. The culmination of our findings are given in Tables 1 and 2.

Overriding the decisions post-hoc provided duplicate results, as the decisions were flipped as per a hard constraint regardless of the learned target values. Although this resulted in significantly lower performance as compared to the baselines, we expect that this technique provides the highest ethical competence. Another benefit of this technique to incorporate morality into an existing learning model is that it is model-agnostic. The ease of applicability of this approach is high, and the complexity is entirely dependent on the complexity of the model.

\begin{table*}[t]
\label{table 1}
\centering
\begin{tabular}{l l l l l l}  
\multicolumn{6}{c}{\textbf{Dataset 1: Graduate Admissions}} \\
\hline
Method & Accuracy & Precision & Recall & F1 & ROC\_AUC \\
\hline
Logistic Regression, No Ethical Competence & 89.72  & 90.78  & 93.53  & 92.13  & 88.19  \\
Overriding Logistic Decision with Hard Moral Constraint & 57.86  & 57.86  & 56.62  & 57.23  & 58.10  \\
Naive Bayes, No Ethical Competence & 90.76  & 95.38  & 88.31  & 91.71  & 91.22  \\
Overriding NB with Hard Moral Constraint & 57.86  & 57.86  & 56.62  & 57.23  & 58.10  \\
Random Forest, No Ethical Competence & 95.57  & 94.93  & 98.38  & 96.60  & 94.40  \\
Random Forest, Inclusion of Morally-aligned weights & 79.40  & 87.86  & 78.98  & 83.19  & 79.64  \\
Overriding RF with Hard Moral Constraint & 81.50  & 18.38  & 98.76  & 31.06  & 90.39  \\
Support Vector Machine, No Ethical Competence & 96.30  & 97.71  & 95.95  & 96.82  & 96.43  \\
Overriding SVM with Hard Moral Constraint & 57.86  & 57.86  & 56.62  & 57.23  & 58.10  \\
Deep Neural Network, No Ethical Competence & 98.30  & 98.81  & 98.65  & 98.54  & 98.24  \\
Overriding DNN with Hard Moral Constraint & 57.86  & 57.86  & 56.62  & 57.23  & 58.10  \\
\textbf{Expected Moral Shortfall,} $\theta = 0.05$ & \textbf{97.38}  & \textbf{98.43}  & \textbf{95.50}  & \textbf{97.70}  & \textbf{97.75}  \\
\textbf{Expected Moral Shortfall,} $\theta = 0.08$ & \textbf{94.11} & \textbf{96.23} & \textbf{89.88} & \textbf{94.67} & \textbf{94.94} \\
\textbf{Expected Moral Shortfall,} $\theta = 1.0$ & \textbf{58.16} & \textbf{54.87} & \textbf{28.08} & \textbf{43.85} & \textbf{64.04} \\
\end{tabular}
\caption{Empirical results for different approaches to ethical competence in learning models\\of varying complexity and design, applied to the first dataset.}
\end{table*}

\begin{table*}
\centering
\begin{tabular}{l l l l l l}  
\multicolumn{6}{c}{\textbf{Dataset 2: Loan Approval}} \\
\hline
Method & Accuracy & Precision & Recall & F1 & ROC\_AUC \\
\hline
Logistic Regression, No Ethical Competence & 80.42  & 73.05  & 16.29  & 26.64  & 57.30  \\
Overriding Logistic Decision with Hard Moral Constraint & 69.46  & 30.68  & 31.73  & 31.20  & 55.86  \\
Naive Bayes, No Ethical Competence & 79.30  & 52.08  & 64.18  & 57.50  & 73.85  \\
Overriding NB with Hard Moral Constraint & 69.46  & 30.68  & 31.73  & 31.20  & 55.86  \\
Random Forest, No Ethical Competence & 93.11  & 96.62  & 70.93  & 81.81  & 85.12  \\
Random Forest, Inclusion of Morally-aligned weights & 82.17  & 7.02  & 26.92  & 11.14  & 55.74  \\
Overriding RF with Hard Moral Constraint & 81.58  & 18.38  & 72.23  & 31.06  & 90.39  \\
Support Vector Machine, No Ethical Competence & 88.92  & 86.42  & 58.41  & 69.71  & 77.92  \\
Overriding SVM with Hard Moral Constraint & 69.46  & 30.68  & 31.73  & 31.20  & 55.86  \\
Deep Neural Network, No Ethical Competence & 91.94  & 91.27  & 65.03  & 76.07  & 81.73  \\
Overriding DNN with Hard Moral Constraint & 69.46  & 30.68  & 31.73  & 31.20  & 55.86  \\
\textbf{Expected Moral Shortfall,} $\theta = 0.05$ & \textbf{91.03}  & \textbf{93.91}  & \textbf{62.98}  & \textbf{75.40}  & \textbf{80.90}  \\
\textbf{Expected Moral Shortfall,} $\theta = 0.08$ & \textbf{89.01}  & \textbf{90.76}  & \textbf{50.73}  & \textbf{66.80}  & \textbf{75.2}0  \\
\textbf{Expected Moral Shortfall,} $\theta = 1.0$ & \textbf{79.4}6  & \textbf{40.89}  & \textbf{10.41}  & \textbf{18.86}  & \textbf{55.20}  \\
\end{tabular}
\caption{Similar results replicated for the second dataset.}
\end{table*}

Adding penalty terms to the sample weights on the basis of $EJ$ values resulted in a reduction of performance from the baseline Random Forests model with no moral considerations. This is because of its local decision-making which makes it more sensitive to perturbations \cite{louppe2014understanding}. Splitting decisions in individual trees lead to a significant deviation from original results, thus the decrease in performance. Adding sample weights to the instances in the linear, probabilistic, and neural models did not lead to a significant variation in results from the baseline, owing to their global regularization. These near redundant results are not presented for the sake of brevity.  

The DNN baseline provided the highest performance in terms of classification accuracy, and when EMS was applied to its loss function, it retained its performance to a great extent. However, the accuracy begins to decline when the value of $\theta$ is increased. This is intuitive, as a larger percentage of the morally worst cases are being affected by the hard constraints. We can see that when we increase $\theta$ to 1, the results become exceedingly close to that of the Overriding performance for the DNN model. This also validates our theory that overriding would provide the best ethical competence in a model, although it would eliminate any learning of the model.

The DNN model has the highest complexity in terms of time and space used, however, it works best with the EMS metric. Therefore, users may takeaway from these results that when prioritizing high ethical competence and low complexity, any model would suffice owing to overriding decisions which can align moral values directly with the model. However, in high-stakes scenarios where decisions are critical and ethical inclusion is pertinent, one may use a deep learning model with the appropriate parameters for the EMS to incorporate desired levels of ethical competence.

\section{Related Work}

Many measures have been taken to incorporate some semblance of moral reasoning into autonomous decision-making systems. For example, \cite{anderson2018geneth} used inductive logic programming to develop a rule-based system that, by design, was general enough to apply to any domain. This is the aim of our proposed model as well, as can be seen from the experiments above. However, using such a discretized model with no learning aspect reduces complex concepts into features too simple to capture large-scale granularity.

MoralDM \cite{dehghani2008integrated}, uses a reasoning approach by considering analogical ethics, or casuistry \cite{casuistry}, which although a valid ethical direction to resolve dilemmas, becomes overly dependent on large-scale datasets. Our approach, the EMS, can be seen working aptly on both a large-scale and small-scale datasets. Like these, most works in this space fall into a \textit{consequentialist trap}, as it seems to be the most intuitive approach to embedding morality. Maximize utility to get the most moral results. However, it is important to note that not all approaches to moral dilemmas are consequentialist, and not all spaces of applied ethics rely on consequentialist means. Therefore, the inclusion of the weighted tuple $W$, allows for a flexible way to make the model lean towards a specific school of thought, without greatly compromising on model performance.

When it comes to optimizing the model to incorporate ethical competence, modifying the loss function to be minimized seemed to be the most apt approach, as opposed to introducing a penalty term or overriding the decision post-hoc. We considered various modification techniques which have appeared in literature adjacent to fairness, robustness, safety, and other similar objectives for moral cognition. One of these was the reinforcement learning with human feedback (RLHF), which is a solid approach for such a task \cite{chaudhari2024rlhf}. However, using human feedback at the learning stage becomes difficult to scale, especially with continuous subjectivity in the case of morality. Although we too have included human-in-the-loop architecture when handling our datasets, the intervention is static and ends with the preprocessing phase. This reduces intervention costs and can be scaled significantly. 

Another approach we considered was optimization using methods such as Distributed Robust Optimization (DRO) \cite{yang2014distributed} and Kahneman and Tversky’s optimization (KTO) \cite{ethayarajh2024kto}. These techniques are usable and may optimize the parameters as per our direction to the model, however, since we are providing the weights $W$ to the model based on our moral preferences, optimizing the results using such parameters may make the model lean towards a specific objective. And since the model is black box, we would be unable to discern which ethical theory is being favored.

Finally, we also considered the application of fairness optimization techniques to enhance ethical competence such as value-aligned learning \cite{al2024training} and fairness-aware machine learning \cite{dunkelau2019fairness}. Although these are good techniques to maximize fairness, they produce lowered performance when applied to morality. Furthermore, aspects such as fairness and safety are a narrow aspect of morality. By including all three normative ethical theories to some degree, the user of the EMS model can ensure that they prioritize principles that they prefer most. 

\section{Conclusion}
In this work we presented a novel method to incorporate ethical competence in AI systems derived from the financial concept of minimizing Expected Shortfall. We call this metric the Expected Moral Shortfall (EMS) and have done rigorous testing and ablation studies by comparing against baselines with some to no ethical competence. We find that EMS presents a flexible way to apply moral reasoning to any domain where such ethical ambiguity is present, and users may capitalize on the flexibility by including their preferences of ethical schools of thought and capacity to withstand risk. We find that EMS gives us the closest results to the baseline through applying hard ethical constraints to the most morally wrong cases.

We believe that our work is thorough, however, there are some assumptions that we have made, leading to notable limitations. Our model has been tested against two datasets where it was possible to map the features to moral values. However, not all real-world data would allow this possibility, and gauging moral concepts from empirical data, such as from the medical domain or the business domain, may become increasingly difficult. Furthermore, if applied in real-world cases with significant real-world impact, there would be a reliance on an ethicist or a team of ethicists in order to make such decisions about the feature mappings and choice of parametric values ($\alpha, \beta, \gamma, \theta$, etc). 

Regardless of the limitations, we believe that our work has significant societal impact on the deployment of AI systems, especially in critical domains where the stakes are high in terms of human lives and property. In the future, this research may be expanded to larger real-world datasets, with more testing and manipulation of the parameters used to determine further specific tradeoffs such as safety and fairness.  

\bibliography{aaai2026}

@article{brozek2019can,
  title={Can artificial intelligences be moral agents?},
  author={Bro{\.z}ek, Bartosz and Janik, Bartosz},
  journal={New ideas in psychology},
  volume={54},
  pages={101--106},
  year={2019},
  publisher={Elsevier}
}

@article{formosa2021making,
  title={Making moral machines: why we need artificial moral agents},
  author={Formosa, Paul and Ryan, Malcolm},
  journal={AI \& society},
  volume={36},
  number={3},
  pages={839--851},
  year={2021},
  publisher={Springer}
}

@article{vanderelst2018architecture,
  title={An architecture for ethical robots inspired by the simulation theory of cognition},
  author={Vanderelst, Dieter and Winfield, Alan},
  journal={Cognitive Systems Research},
  volume={48},
  pages={56--66},
  year={2018},
  publisher={Elsevier}
}

@article{moor2006nature,
  title={The nature, importance, and difficulty of machine ethics},
  author={Moor, James H},
  journal={IEEE intelligent systems},
  volume={21},
  number={4},
  pages={18--21},
  year={2006},
  publisher={IEEE}
}

@book{spiekermann2023value,
  title={Value-based engineering: a guide to building ethical technology for humanity},
  author={Spiekermann, Sarah},
  year={2023},
  publisher={De Gruyter}
}

@InCollection{sep-morality-definition,
	author       =	{Gert, Bernard and Gert, Joshua},
	title        =	{{The Definition of Morality}},
	booktitle    =	{The {Stanford} Encyclopedia of Philosophy},
	editor       =	{Edward N. Zalta},
	howpublished =	{\url{https://plato.stanford.edu/archives/fall2020/entries/morality-definition/}},
	year         =	{2020},
	edition      =	{{F}all 2020},
	publisher    =	{Metaphysics Research Lab, Stanford University}
}

@misc{aijaz2025appleappliedethicsontology,
      title={ApplE: An Applied Ethics Ontology with Event Context}, 
      author={Aisha Aijaz and Raghava Mutharaju and Manohar Kumar},
      year={2025},
      eprint={2502.05110},
      archivePrefix={arXiv},
      primaryClass={cs.CY},
      url={https://arxiv.org/abs/2502.05110}, 
}

@misc{casuistry,
    title={Casuistry},
    author={D. P. Schmidt},
    year={2023},
    publisher={Encyclopedia Britannica},
    url={https://www.britannica.com/topic/casuistry}
}

@article{gordon2020building,
  title={Building moral robots: Ethical pitfalls and challenges},
  author={Gordon, John-Stewart},
  journal={Science and engineering ethics},
  volume={26},
  number={1},
  pages={141--157},
  year={2020},
  publisher={Springer}
}

@article{cervantes2020toward,
  title={Toward ethical cognitive architectures for the development of artificial moral agents},
  author={Cervantes, Salvador and L{\'o}pez, Sonia and Cervantes, Jos{\'e}-Antonio},
  journal={Cognitive systems research},
  volume={64},
  pages={117--125},
  year={2020},
  publisher={Elsevier}
}

@book{kagan2018normative,
  title={Normative ethics},
  author={Kagan, Shelly},
  year={2018},
  publisher={Routledge}
}

@article{awad2018moral,
  title={The moral machine experiment},
  author={Awad, Edmond and Dsouza, Sohan and Kim, Richard and Schulz, Jonathan and Henrich, Joseph and Shariff, Azim and Bonnefon, Jean-Fran{\c{c}}ois and Rahwan, Iyad},
  journal={Nature},
  volume={563},
  number={7729},
  pages={59--64},
  year={2018},
  publisher={Nature Publishing Group}
}

@article{anderson2018geneth,
  title={GenEth: A general ethical dilemma analyzer},
  author={Anderson, Michael and Anderson, Susan Leigh},
  journal={Paladyn, Journal of Behavioral Robotics},
  volume={9},
  number={1},
  pages={337--357},
  year={2018},
  publisher={Sciendo}
}

@inproceedings{greene2016embedding,
  title={Embedding ethical principles in collective decision support systems},
  author={Greene, Joshua and Rossi, Francesca and Tasioulas, John and Venable, Kristen and Williams, Brian},
  booktitle={Proceedings of the AAAI conference on artificial intelligence},
  volume={30},
  year={2016}
}

@book{bergmann2014challenges,
  title={Challenges to Moral and Religious Belief: Disagreement and Evolution},
  author={Bergmann, M. and Kain, P.},
  isbn={9780191648540},
  year={2014},
  publisher={OUP Oxford}
}

@article{tasche2002expected,
  title={Expected shortfall and beyond},
  author={Tasche, Dirk},
  journal={Journal of Banking \& Finance},
  volume={26},
  number={7},
  pages={1519--1533},
  year={2002},
  publisher={Elsevier}
}

@article{whitby2003myth,
  title={The Myth of AI Failure CSRP 568},
  author={Whitby, Blay},
  journal={Cognitive Science Research Paper-University Of Sussex CSRP},
  year={2003},
  publisher={University Of Sussex}
}

@book{singer1986applied,
  title={Applied ethics},
  author={Singer, Peter},
  year={1986},
  publisher={Oxford University Press}
}

@article{aijaz2025moralcompass,
    author = {Aijaz, Aisha and Batra, Arnav and Bazaaz, Aryaan and Srinivasa, Srinath and Mutharaju, Raghava and Kumar, Manohar},
    title = {Moral Compass: A Data-Driven Benchmark for Ethical Cognition in AI},
    journal = {34th International Joint Conference on Artificial Intelligence},
    year = {2025}
}

@article{loanapprovaldataset,
    author = {Kumar, Akshay and Andrey, Martynov and S. Barik, Dibya and R., Vishwa},
    title = {Loan Approval Prediction},
    journal = {Kaggle Datasets},
    year = {2024}
}

@article{gradadmissionsdataset,
    author = {Manral, Mukesh},
    title = {Graduates Admission Prediction},
    journal = {Kaggle Datasets},
    year = {2022}
}

@article{bleske2014trends,
  title={Trends in GRE scores and graduate enrollments by gender and ethnicity},
  author={Bleske-Rechek, April and Browne, Kingsley},
  journal={Intelligence},
  volume={46},
  pages={25--34},
  year={2014},
  publisher={Elsevier}
}

@article{wait2009relationship,
  title={Relationship between TOEFL score and academic success for international engineering students},
  author={Wait, Isaac W and Gressel, Justin W},
  journal={Journal of engineering education},
  volume={98},
  number={4},
  pages={389--398},
  year={2009},
  publisher={Wiley Online Library}
}

@article{sowunmioptimization,
  title={OPTIMIZATION IN FINANCE},
  author={Sowunmi, BSc Ololade and Popela, RNDr Pavel},
  journal={Brno University of Technology},
  year={2020}
}

@article{rockafellar2000optimization,
  title={Optimization of conditional value-at-risk},
  author={Rockafellar, R Tyrrell and Uryasev, Stanislav and others},
  journal={Journal of risk},
  volume={2},
  pages={21--42},
  year={2000}
}

@article{zhang2025research,
  title={Research on the Impact of Different Measurements on Admission in Higher Education},
  author={Zhang, Boyang},
  journal={Science and Technology of Engineering, Chemistry and Environmental Protection},
  volume={1},
  number={3},
  year={2025}
}

@article{dunkelau2019fairness,
  title={Fairness-aware machine learning},
  author={Dunkelau, Jannik and Leuschel, Michael},
  journal={An extensive overview},
  pages={1--60},
  year={2019}
}

@article{al2024training,
  title={Training value-aligned reinforcement learning agents using a normative prior},
  author={Al Nahian, Md Sultan and Frazier, Spencer and Riedl, Mark and Harrison, Brent},
  journal={IEEE Transactions on Artificial Intelligence},
  volume={5},
  number={7},
  pages={3350--3361},
  year={2024},
  publisher={IEEE}
}

@article{ethayarajh2024kto,
  title={Kto: Model alignment as prospect theoretic optimization},
  author={Ethayarajh, Kawin and Xu, Winnie and Muennighoff, Niklas and Jurafsky, Dan and Kiela, Douwe},
  journal={arXiv preprint arXiv:2402.01306},
  year={2024}
}

@article{yang2014distributed,
  title={Distributed robust optimization (DRO), part I: Framework and example},
  author={Yang, Kai and Huang, Jianwei and Wu, Yihong and Wang, Xiaodong and Chiang, Mung},
  journal={Optimization and Engineering},
  volume={15},
  number={1},
  pages={35--67},
  year={2014},
  publisher={Springer}
}

@article{chaudhari2024rlhf,
  title={Rlhf deciphered: A critical analysis of reinforcement learning from human feedback for llms},
  author={Chaudhari, Shreyas and Aggarwal, Pranjal and Murahari, Vishvak and Rajpurohit, Tanmay and Kalyan, Ashwin and Narasimhan, Karthik and Deshpande, Ameet and Castro da Silva, Bruno},
  journal={ACM Computing Surveys},
  year={2024},
  publisher={ACM New York, NY}
}

@article{krithikadatta2014normal,
  title={Normal distribution},
  author={Krithikadatta, Jogikalmat},
  journal={Journal of Conservative Dentistry and Endodontics},
  volume={17},
  number={1},
  pages={96--97},
  year={2014},
  publisher={Medknow}
}

@article{louppe2014understanding,
  title={Understanding random forests},
  author={Louppe, Gilles},
  journal={Cornell University Library},
  volume={10},
  year={2014}
}

@article{ruby2020binary,
  title={Binary cross entropy with deep learning technique for image classification},
  author={Ruby, Usha and Yendapalli, Vamsidhar and others},
  journal={Int. J. Adv. Trends Comput. Sci. Eng},
  volume={9},
  number={10},
  year={2020}
}

@book{heath2014morality,
  title={Morality, competition, and the firm: The market failures approach to business ethics},
  author={Heath, Joseph},
  year={2014},
  publisher={Oxford University Press}
}

@book{holmes2018introduction,
  title={Introduction to Applied Ethics},
  author={Holmes, Robert L},
  year={2018},
  publisher={Bloomsbury Publishing}
}

@article{sheetz2020trends,
  title={Trends in the adoption of robotic surgery for common surgical procedures},
  author={Sheetz, Kyle H and Claflin, Jake and Dimick, Justin B},
  journal={JAMA network open},
  volume={3},
  number={1},
  pages={e1918911--e1918911},
  year={2020},
  publisher={American Medical Association}
}

@article{cominelli2018seai,
  title={SEAI: Social emotional artificial intelligence based on Damasio’s theory of mind},
  author={Cominelli, Lorenzo and Mazzei, Daniele and De Rossi, Danilo Emilio},
  journal={Frontiers in Robotics and AI},
  volume={5},
  pages={6},
  year={2018},
  publisher={Frontiers Media SA}
}

@inproceedings{abdullah2022chatgpt,
  title={ChatGPT: Fundamentals, applications and social impacts},
  author={Abdullah, Malak and Madain, Alia and Jararweh, Yaser},
  booktitle={2022 Ninth International Conference on Social Networks Analysis, Management and Security (SNAMS)},
  pages={1--8},
  year={2022},
  organization={Ieee}
}

@inproceedings{dehghani2008integrated,
  title={An Integrated Reasoning Approach to Moral Decision-Making.},
  author={Dehghani, Morteza and Tomai, Emmett and Forbus, Kenneth D and Klenk, Matthew},
  booktitle={AAAI},
  pages={1280--1286},
  year={2008}
}

@inproceedings{briggs2020mitigating,
  title={Mitigating discrimination in clinical machine learning decision support using algorithmic processing techniques},
  author={Briggs, Emma and Hollm{\'e}n, Jaakko},
  booktitle={Discovery Science: 23rd International Conference, DS 2020, Thessaloniki, Greece, October 19--21, 2020, Proceedings 23},
  pages={19--33},
  year={2020},
  organization={Springer}
}

@article{10.1093/jlb/lsac012,
    author = {Papalexopoulos, Theodore P and Bertsimas, Dimitris and Cohen, I Glenn and Goff, Rebecca R and Stewart, Darren E and Trichakis, Nikolaos},
    title = "{Ethics-by-design: efficient, fair and inclusive resource allocation using machine learning}",
    journal = {Journal of Law and the Biosciences},
    volume = {9},
    number = {1},
    pages = {lsac012},
    year = {2022},
    month = {04},
    issn = {2053-9711},
    doi = {10.1093/jlb/lsac012},
    url = {https://doi.org/10.1093/jlb/lsac012},
    eprint = {https://academic.oup.com/jlb/article-pdf/9/1/lsac012/43492134/lsac012.pdf},
}

@article{demarco1997coherence,
  title={Coherence and applied ethics},
  author={DeMarco, Joseph P},
  journal={Journal of applied philosophy},
  volume={14},
  number={3},
  pages={289--300},
  year={1997},
  publisher={Wiley Online Library}
}

@article{van2002deontic,
  title={Deontic Logic and Computer-Supported Computer Ethics},
  author={Van Den Hoven, Jeroen and Lokhorst, Gert-Jan},
  journal={Metaphilosophy},
  volume={33},
  number={3},
  pages={376--386},
  year={2002},
  publisher={Wiley Online Library}
}

\end{document}